\documentclass[aps,nofootinbib,twocolumn,floatfix]{revtex4}
\usepackage{amsmath, amsthm, amssymb,graphicx}
\begin{document}

\newcommand{\rrangle}{\rangle\!\rangle} \newcommand{\llangle}{\langle\!\langle}

\newcommand{\cM}{\mathcal{M}}
\newcommand{\cG}{\mathcal{G}}
\newcommand{\cD}{\mathcal{D}}
\newcommand{\cE}{\mathcal{E}}
\newcommand{\cL}{\mathcal{L}}
\newcommand{\cU}{\mathcal{U}}
\newcommand{\cH}{\mathcal{H}}

\newcommand{\reals}{\mathbb{R}}

\newcommand{\diff}{\mathrm{d}\!}
\newcommand{\pdiff}[2]{\frac{\partial #1}{\partial #2}}
\newcommand{\pdiffsq}[3]{\frac{\partial^2 #1}{\partial #2\partial #3}}
\newcommand{\expect}[1]{\ensuremath{\left\langle#1\right\rangle}}

\newcommand{\qhat}{\hat{q}}
\newcommand{\rhohat}{\hat{\rho}}
\newcommand{\rhoMLE}{\hat{\rho}_{\mathrm{MLE}}}
\newcommand{\rhotrue}{\rho_{\mathrm{true}}}
\newcommand{\Rhat}{\hat{\mathcal{R}}}
\newcommand{\Vbar}{\overline{V}}

\newcommand{\ket}[1]{\ensuremath{\left|#1\right\rangle}}
\newcommand{\bra}[1]{\ensuremath{\left\langle#1\right|}}
\newcommand{\braket}[2]{\ensuremath{\left\langle#1|#2\right\rangle}}

\newcommand{\ketbra}[2]{\ket{#1}\!\!\bra{#2}}
\newcommand{\braopket}[3]{\ensuremath{\bra{#1}#2\ket{#3}}}
\newcommand{\proj}[1]{\ketbra{#1}{#1}}

\newcommand{\sket}[1]{\ensuremath{\left|#1\right\rrangle}}
\newcommand{\sbra}[1]{\ensuremath{\left\llangle#1\right|}}
\newcommand{\sbraket}[2]{\ensuremath{\left\llangle#1|#2\right\rrangle}}
\newcommand{\sketbra}[2]{\sket{#1}\!\!\sbra{#2}}
\newcommand{\sbraopket}[3]{\ensuremath{\sbra{#1}#2\sket{#3}}}
\newcommand{\sproj}[1]{\sketbra{#1}{#1}}

\def\Id{1\!\mathrm{l}}
\newcommand{\Tr}{\mathrm{Tr}}
\newcommand{\Nparams}{N_{\mathrm{params}}}

\title{Robust, self-consistent, closed-form tomography of quantum logic gates on a trapped ion qubit}
\begin{abstract}
We introduce and demonstrate experimentally: (1) a framework called ``gate set tomography'' (GST) for self-consistently characterizing an entire set of quantum logic gates on a black-box quantum device; (2) an explicit closed-form protocol for linear-inversion gate set tomography (LGST), whose reliability is independent of pathologies such as local maxima of the likelihood; and (3) a simple protocol for objectively scoring the accuracy of a tomographic estimate without reference to target gates, based on how well it predicts a set of testing experiments.  We use gate set tomography to characterize a set of Clifford-generating gates on a single trapped-ion qubit, and compare the performance of (i) standard process tomography; (ii) linear gate set tomography; and (iii) maximum likelihood gate set tomography.
\end{abstract}

\author{Robin Blume-Kohout}
\author{John King Gamble}
\author{Erik Nielsen}
\author{Jonathan Mizrahi}
\author{Jonathan D. Sterk}
\author{Peter Maunz}
\affiliation{Sandia National Laboratories, Albuquerque, New Mexico 87185}
\date{\today}
\maketitle

Quantum information processing (QIP) relies upon precise, repeatable quantum logic operations.  Experiments in multiple QIP technologies \cite{MerkelPRA13,TakahashiPRA13,MedfordNN13,MahlerPRL13,BrownPRA11} have implemented quantum logic gates with sufficient precision to reveal weaknesses in the \emph{quantum tomography} protocols used to characterize those gates.  Conventional tomographic methods assume and rely upon a precalibrated reference frame, comprising (1) the measurements performed on unknown states, and (2) for quantum process tomography, the test states that are prepared and fed into the process (gate) to be characterized.  Standard process tomography on a gate $G$ proceeds by repeating a series of experiments in which state $\rho_j$ is prepared and observable (a.k.a. \emph{POVM effect}) $E_k$ is observed, using the statistics of each such experiment to estimate the corresponding probability
$$p_{k|j} = \Tr[ E_k G[\rho_j] ]$$
(given by Born's rule), and finally reconstructing $G$ from many such probabilities.  


\begin{figure}[t]
\includegraphics[width=3in]{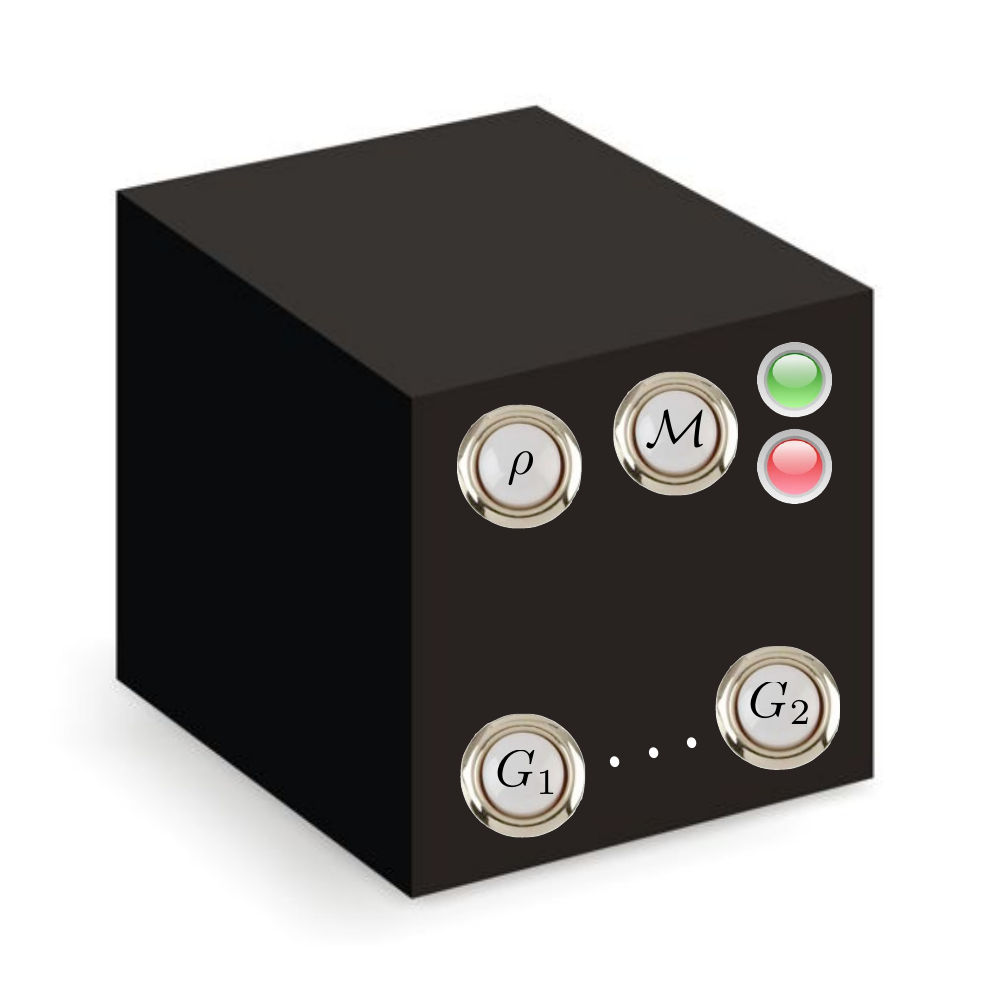}
\caption{\textbf{The GST model of a quantum device.}  Gate set tomography treats the quantum system of interest as a black box, with strictly limited access.  This is a fairly good model for many qubit technologies, especially those based on solid state and/or cryogenic technologies.  We do not have direct access to the Hilbert space or any aspect of it.  Instead, the device is controlled via \emph{buttons} that implement various gates (including a preparation gate and a measurement that causes one of two indicator lights to illuminate).  Prior information about the gates' function may be available, and can be used, but should not be \emph{relied} upon.} \label{fig:blackbox}
\end{figure}

But, in most QIP technologies, the various test states ($\rho_j$) and measurement outcomes ($E_k$) are \emph{not} known exactly.  Instead, they are implemented using the very same gates that process tomography is supposed to characterize.  The quantum device is effectively a black box, accessible only via classical control and classical outcomes of quantum measurements, and in this scenario standard tomography can be dangerously self-referential.  If we do process tomography on gate $G$ under the common assumption that the test states and measurement outcomes are both eigenstates of the Pauli $\sigma_x, \sigma_y, \sigma_z$ operators, then the accuracy of the estimate $\hat{G}$ will be limited by the error in this assumption.

This is now a critical experimental issue.  In platforms including (but not limited to) superconducting flux qubits \cite{MerkelPRA13}, trapped ions \cite{BrownPRA11}, and solid-state qubits, quantum logic gates are being implemented so precisely that systematic errors in tomography (due to miscalibrated reference frames) are glaringly obvious.  Fixes have been proposed \cite{MogilevtsevNJP12,BranczykNJP12,MerkelPRA13,TakahashiPRA13}, but none yet provide a general, comprehensive, reliable scheme for gate characterization.  In this article, we present \emph{gate set tomography} (GST), a complete scheme for reliably and accurately characterizing an entire set of quantum gates.  In particular we introduce the first \emph{linear-inversion} protocol for self-consistent process tomography, linear gate set tomography (LGST).  LGST is a closed-form estimation protocol (inspired in part by \cite{Stark12a,Stark12b,Stark12c}) that cannot -- unlike pure maximum-likelihood (ML) algorithms -- run afoul of local maxima in a likelihood function that is generally ill-behaved.  While the price of LGST's reliability is decreased accuracy compared with ML, it is easy to recover accuracy using a hybrid scheme in which the LGST estimate is used as the starting point for local ML estimation.  We demonstrate (L)GST experimentally by characterizing a complete set of gates for a trapped-ion qubit.  To demonstrate its performance, we introduce a novel quantitative scoring protocol that evaluates how well a tomographic estimate predicts independent ``test'' experiments.

\section{Background} \label{sec:Background}

We begin with a brief review of standard tomography, and the mathematical conventions used in both standard and gate set tomography.
 
\subsection{Mathematical conventions}

A quantum information processing device is described using a Hilbert space $\cH$ of finite dimension $d$ (any system with $d=2$ is a \emph{qubit}).  Its state at any time is described by a $d\times d$ \emph{density matrix} $\rho$ that is positive semidefinite ($\rho\geq0$) and has unit trace ($\Tr\rho=1$).  Each possible measurement on the system is represented by a \emph{positive, operator-valued measure} (POVM), comprising a set $\cM = \{E_k\}$ of $d\times d$ matrices that are positive semidefinite ($E_k\geq0$) and sum to the identity matrix ($\sum_k{E_k} = \Id$).  This representation gains its meaning from Born's rule, which states that when measurement $\cM$ is performed on a system in state $\rho$, outcome ``$k$'' will be observed with probability
$$Pr(k|\rho) = \Tr[E_k\rho].$$
A system's reversible evolution is described by some $d\times d$ unitary operator $U$, and the state evolves as $\rho_{t_2} = U\rho_{t_1} U^\dagger$.  In practice, a system's dynamics (e.g., when a logic gate is applied in the laboratory) will be at least slightly irreversible, and must be represented by a \emph{completely positive, trace-preserving linear map on density matrices} (CPTP map) that can be written in Kraus form:
$$G[\rho] = \sum_i{K_i \rho K_i^\dagger},$$
where the $\{K_i\}$ are matrices satisfying $\sum_i{K_i K_i^\dagger}=\Id$.  

In tomography, it is more useful to represent quantum processes using the \emph{Hilbert-Schmidt space} of matrices on $\cH$, denoted $B(\cH)$, in which any $d\times d$ matrix $X$ is a column vector $\sket{X}$ or row vector $\sbra{X}$.  In this representation, we can write Born's Rule as
\begin{equation}
Pr(k|\rho) = \sbraket{E_k}{\rho}
\end{equation}
using the Hilbert-Schmidt inner product $\sbraket{X}{Y} \equiv \Tr[X^\dagger Y]$.  Since quantum processes are always linear maps on density matrices, they can always be represented as $d^2\times d^2$ matrices (a.k.a. superoperators) on Hilbert-Schmidt space.  In this representation, if a process $G$ is applied to a state $\rho$ and then a measurement $\cM = \{E_k\}$ is performed, then the probability of observing outcome ``$k$'' is simply
$$Pr(k|\rho,G) = \sbraopket{E_k}{G}{\rho}.$$

\subsection{Standard tomography}

In this framework, quantum state tomography \cite{VogelPRA89,SmitheyPRL93,ParisBook04} is a simple linear algebraic inversion.  Given an unknown state $\sket{\rho}$, we characterize it by performing a set of measurements that are \emph{informationally complete} -- i.e., their outcomes $\{E_k\}$ collectively span $B(\cH)$.  We repeat the measurement(s) $N$ times, count the observations of outcome ``$k$'' ($n_k$), estimate its probability as
\begin{equation}
Pr(k|\rho) \approx \frac{n_k}{N} \equiv \hat{p}_k,
\end{equation}
and use linear algebra to invert the set of equations
\begin{equation}
\sbraket{E_k}{\rho} = \hat{p}_k.
\end{equation}

Quantum process tomography \cite{ChuangJMO97,ParisBook04} is very similar, but in addition to an informationally complete set of measurement outcomes, we must also prepare a set of test states $\rho_j$ that span $B(\cH)$, and apply the unknown process $G$ to them.  From the count statistics of these repeated experiments, we estimate
\begin{equation}
Pr(k|G,\rho_j) \approx \frac{n_{j,k}}{N} \equiv \hat{p}_{j,k},
\end{equation}
and use linear algebra to invert the set of equations
\begin{equation}
\sbraopket{E_k}{G}{\rho_j} = \hat{p}_{j,k}.
\end{equation}

These techniques define \emph{linear-inversion tomography}.  While they have been largely supplanted by more sophisticated statistical methods \cite{HradilPRA97,RBKNJP10,RBKPRL10} that provide better accuracy for finite $N$, linear-inversion techniques remain useful in the limit $N\to\infty$.  More importantly, their existence guarantees the existence of efficient, reasonably accurate protocols for standard tomography.  They can serve as a first stage in more accurate protocols \cite{SmolinPRL12}, or enable existence proofs for more sophisticated protocols \cite{CramerNC10,GrossPRL10}.

These protocols, however, assume that the $\{\rho_j\}$ and $\{E_k\}$ are known -- and use them to define an absolute reference frame for the quantum system's state space.  This is a theorist's fiction; in every quantum technology except (arguably) linear optics, no such reference frame is given.  The assortment of test states and measurements required for process tomography are obtained by applying the same dynamical gates that process tomography is supposed to characterize to a \emph{single} imperfectly known fiducial state $\rho$ and a single imperfectly known fiducial measurement $\cM$.

\section{Gate set tomography} \label{sec:GST}

Gate set tomography (GST) is based on a simple insight:  playing around with a quantum device should be sufficient to reveal all the properties needed to predict its future behavior.  If this is true, then every unjustified assumption should also be unnecessary.  We need not -- and should not -- assert that certain operations prepare $\ket{0}$ states, measure the $\sigma_z$ basis, etc.  If they do operate this way, then the data will reveal it.

To enforce this intellectual discipline, we model the quantum device as a black box (see Fig. \ref{fig:blackbox}, and also Ref. \cite{vanDamACM00}, which pioneered the idea that black box qubits should be fully characterizable).  Our interaction with the black box is strictly classical and limited to pushing a small number of ``buttons'' (generally implemented in experiments by electromagnetic control pulses):
\begin{itemize}
\item One button, marked ``$\rho$'', initializes the system.
\item Another button, marked ``$\cM$'', performs a 2-outcome measurement -- it is accompanied by 2 lights, exactly one of which lights up to indicate the outcome.
\item Finally, a set of $K$ buttons labeled $G_1\ldots G_K$ perform quantum operations (logic gates) on the system.
\end{itemize}

All of these buttons' effects are unknown, and have to be deduced from the data.  No other controls exist.  In this article, we will make a number of simplifying assumptions, all of which can be relaxed (at some cost -- which will be discussed further in \cite{RBKinprep}) to make GST more robust.
\begin{itemize}
\item The Hilbert space dimension $d=\mathrm{dim}(\cH)$ is known.
\item The effect of the initialization button really is to reprepare the system (repeatably) in a state $\rho$.
\item The measurement button $\cM$ can be represented by a 2-outcome POVM $\{E,\Id-E\}$.
\item Control is Markovian:  Each button can be represented by a completely positive, trace-preserving (CPTP) map on $B(\cH)$.
\end{itemize}
A \emph{gate set}, then, is a complete description of a black box.  In Hilbert-Schmidt space notation, it is
$$\cG = \{\sket{\rho},\sbra{E},\{G_k\}\}.$$
The goal of GST is to identify $\cG$ from the results of experiments on the black box system.

\subsection{Experiments, data, and inference in GST}

GST, like all tomography protocols, comprises (1) obtaining data, and (2) analyzing the data to get an estimate.  Data are gathered from a discrete set of $M$ \emph{experiments}, each of which is repeated many ($N$) times to get statistics.  Experiments have a simple form:
\begin{enumerate}
\item Push the ``$\rho$'' button to initialize the system.
\item Apply a sequence $s = \{G_{s_1}, G_{s_2}, G_{s_3},\ldots,G_{s_L}\}$ of $L$ gates.
\item Push the ``$\cM$'' button and record the outcome.
\end{enumerate}
Experiments are described and indexed by the sequence $s$, and the data comprise the observed counts $\{n_s\}$ (for each of the $M$ values of $s$ that were performed).  Note that sequences of gates are equivalent to quantum circuits, except that in circuit design it is usually assumed that certain gates commute (i.e., gates on different qubits) and can be performed in parallel.  Since this may be violated in experimental hardware, we do not assume it in GST -- if two gates \emph{do} commute, it will be apparent in the data.  Since sequences correspond to circuits, we can see GST as predicting the statistics of \emph{arbitrary} circuits by studying the behavior of a \emph{specific} (and limited) set of circuits.

Each experiment $s$ has two outcomes, and is thus associated with a single probability
$$p_s = Pr(E|\rho,s) = \sbraopket{E}{G_{s_L}\circ G_{s_{L-1}}\circ\ldots\circ G_{s_2}\circ G_{s_1}}{\rho}.$$
That experiment's observed counts ($n_s$) provide information about $p_s$, which is a single parameter of the gate set $\cG$.  A simple if crude inference procedure is to estimate
\begin{equation}
\hat{p}_s = \frac{n_s}{N},
\end{equation}
and thus to nail down the parameters of $\cG$ one by one.  If $\mathrm{dim}(\cH)=d$, then $\rho$ requires $d^2-1$ parameters, $E$ requires $d^2$, and each of the $K$ gates $G_k$ requires $d^2(d^2-1)$, suggesting that
$$M \approx Kd^4-(K-2)d^2-1$$
distinct experiments should be necessary and sufficient to identify $\cG$.  

\subsection{The gauge} \label{sec:Gauge}

Not every parameter in $\cG = \{\rho,E,\{G_k\}\}$ can be estimated, however.  Given a gate set $\cG$, let $M$ be some invertible $d^2\times d^2$ superoperator, and suppose that we construct a different gate set $\cG'$ given by
\begin{eqnarray}
\sket{\rho'} &=& M\sket{\rho} \nonumber \\
\sbra{E'} &=& \sbra{E}M^{-1} \nonumber \\
G'_k &=& M G_k M^{-1}.
\end{eqnarray}
Every observable probability $p_s = \sbraopket{E}{G_{s_L}\circ \ldots\circ G_{s_1}}{\rho}$ is identical for $\cG$ and $\cG'$.  So the action of $M$ is a \emph{gauge transformation} (see also \cite{MerkelPRA13}), and $\cG'$ and $\cG$ are equivalent representations of the same physical gate set.  The gauge group is $SL(d^2)$, since $M$ must be invertible, and $M$ and $\alpha M$ act identically for any scalar $\alpha$.

This gauge freedom means that the standard representation of a gate set as $\{\rho,E,\{G_k\}\}$ (which coincides with the way that operations and states are conventionally represented in quantum information) contains approximately $d^2-1$ redundant and unobservable parameters.  Instead of describing the system's observable dynamics, they define only the convenient but arbitrary reference frame (akin to the conventional $\hat{x},\hat{y},\hat{z}$ axes in space) in which a given experimentalist or theorist has chosen to express those dynamics.  For example,
$$\cG_0 = \{\rho=E=\proj{0},G_1=e^{i\sigma_z\pi/4},G_2=e^{i\sigma_x\pi/4}\}$$
is indistiguishable from
$$\hat{\cG} = \{\rho=E=\proj{+},G_1=e^{i\sigma_x\pi/4},G_2=e^{i\sigma_y\pi/4}\}.$$
The only difference is what we have chosen to call the computational basis.

We do not yet know any satisfying gauge-invariant ``normal form'' for gate sets (although some of the gauge parameters can be fixed in obvious ways, e.g. by defining $\rho$ to be diagonal in the computational basis), nor a well-motivated gauge-invariant measure of fidelity between gate sets.  The sets $\cG_0$ and $\hat{\cG}$ shown above appear quite different, and the gate-by-gate fidelity between them would be very low by any measure.  Yet they are in fact indistinguishable.  If an experimentalist set out to implement $\cG_0$, it would be quite unfair if a tomographer reported that (1) in fact $\hat{\cG}$ was being implemented, and therefore (2) the fidelity of implementation is quite low!  Ultimately, we aspire to a gauge-invariant theory, or at least a canonical way of fixing the gauge.  In its absence, we compare a tomographic estimate $\hat{\cG}$ to a target $\cG_0$ by optimizing the gauge numerically.

To compare $\hat{\cG}$ with $\cG_0$, we search for the gauge transformation $M\in SL(d^2)$ that makes $\hat{\cG}$ as similar as possible to $\cG_0$ (e.g., as measured by $\sum_k{||G_k-\hat{G}_k||_2^2}$).  Obviously, scientific integrity suggests that the intended target should play no role in the estimation of physically observable quantities.  So first, we perform tomography and obtain an estimate $\hat{\cG}$ without considering the gauge (or the target).  Only then do we vary the gauge in which $\hat{\cG}$ is described (which has no effect on anything observable) to minimize the discrepancy between $\cG_0$ and $\hat{\cG}$.

\subsection{Positivity}

Positivity is a highly desirable property of any theory; it means that no matter what weird objects appear internal to the theory, every observable probability is always in the range $[0,1]$.  In the conventional representation of quantum operations, positivity demands that:
\begin{itemize}
\item $\rho$ is a positive semidefinite operator with trace 1,
\item $E$ and $\Id-E$ are positive semidefinite,
\item Each $G_k$ is a CPTP map [i.e., $(G_k\otimes\Id)[\rho]\geq0\ \forall\ \rho\geq0$].
\end{itemize}
These conditions are always sufficient for positivity, but not strictly necessary in the black box model.  In the black box model, we cannot prepare arbitrary states (so $E$ need not be strictly positive), nor perform arbitrary measurements (so $\rho$ need not be strictly positive), nor inject systems that are entangled with external ancillae (so complete positivity is something of a red herring).  But since we reasonably anticipate that quantum mechanics is the same inside the black box as outside, it is reasonable to demand that our estimate satisfy the conventional positivity conditions anyway.

However, gauge transformations do not respect the conventional positivity constraints.  Gauge transformation of a gate set in which every gate is CPTP can easily yield a gate set in which several (if not all) of the gates violate complete positivity.  It is natural to define a CPTP gate set as one that is gauge-equivalent to a set of CPTP gates, but we have no closed-form test for this property.  An alternative is to demand that each gate $G_k$ be explicitly CP, but this constraint truncates the gauge freedom.  For an extremal gate set -- where $\rho$ and $E$ are rank-1 projectors, and each $G_k$ is a unitary -- complete positivity simply reduces the gauge group to $SU(d)$.  Every gauge transformation that does not lie in this subgroup would produce a new gate set that violated positivity constraints.

For noisy gate sets, in which each gate lies in the interior of the set of CPTP maps, requiring positivity has more complicated consequences for the gauge.  Any sufficiently small $SL(d^2)$ transformation will preserve positivity.  But if a gauge transformation outside of the $SU(d)$ subgroup is iterated enough times, then it will eventually violate positivity.  Gauge transformations do not form a group.  There may exist pairs of CPTP gate sets that are gauge-equivalent, yet are not connected by a continuous path of gauge transformations.

These complications are severe enough that in this work, we do \emph{not} impose complete positivity, much as as early work on linear-inversion state tomography did not impose positive semidefiniteness on the density matrix $\hat{\rho}$.  Instead, we allow the estimated gates to be arbitrary matrices, and rely on consistency with experimental data to ensure that the gate sets predict positive probabilities for future experiments.  Careful implementation of positivity constraints is a clear and pressing subject for future work.

\subsection{Practical inference of the gate set}

Estimating $\hat{p}_s = \frac{n_s}{N}$ and reconstructing $\cG$ from these estimated probabilities is impractical.  For one thing, the linear inversion estimate of $p_s$ is imperfect ($n_s$ is rarely equal to $p_sN$).  A more elegant and robust approach is to define a \emph{likelihood function} over gate sets,
\begin{equation}
\cL(\cG) = \prod_s{p_s(\cG)^{n_s} (1-p_s(\cG))^{N-n_s}},
\end{equation}
and construct an estimate from it.  A simple and popular (albeit still suboptimal; see discussion in \cite{RBKNJP10}) technique is maximum likelihood estimation, which reports
$$\hat{\cG}_{ML} = \mathrm{argmax}[\cL(\cG)].$$
But here, GST (and other techniques for self-consistent gate estimation) diverge from standard state and process tomography.  In standard tomography the likelihood function is log-convex \cite{RBKNJP10}, and therefore has a unique local maximum that can be found via a variety of numerical techniques.  But in GST, the observable probabilities $p_s(\cG)$ are not linear functions of the parameters of $\cG$.  They are polynomials of degree $L$, because each gate can appear up to $L$ times in an experiment (e.g., $p = \sbraopket{E}{G_1^L}{\rho}$).  This makes the crude approach of estimating the probabilities $\{\hat{p}_s\}$ and then solving for $\cG$ almost impossible, but (more worryingly), it also means that the GST likelihood function will not generally be log-convex or have a unique local maximum.

Prior approaches made use of the assumption that the target gates $\cG_0$ are a good prior estimate of $\cG$, e.g. by maximizing a series expansion of $\cL(\cG)$ in the neighborhood of $\cG_0$ \cite{MerkelPRA13}.  This may work in many cases, but it depends critically on the accuracy of the prior knowledge -- and could lead to worryingly circular estimates (i.e., MLE may find a local maximum near the prior estimate, even if the prior estimate is wildly wrong and the true global maximum is far away).  In the next section, we solve this problem with a robust, closed-form estimator that can be used directly, or as a reliable ``pretty close'' starting point for MLE.

\section{Linear gate set tomography} \label{sec:LGST}

Linear inversion state tomography, the oldest and simplest form of tomography \cite{VogelPRA89,ChuangJMO97}, is based on the notion that we should assign an estimate $\rhohat$ that predicts probabilities equal to observed frequencies:
\begin{equation}
\Tr(\rhohat E) = Pr(E|\rhohat) = \frac{n_E}{N}. \label{eq:LI}
\end{equation}
When such a $\rhohat$ (1) exists and (2) is physically valid, it will also maximize the likelihood.  So linear inversion and MLE coincide in such cases.  When the data are overcomplete, the set of equations implied by Eq. \ref{eq:LI} are overconstrained.  Linear inversion is still possible using least-squares inversion, which minimizes a weighted sum of residuals between probabilities and observed frequencies,
\begin{equation}
\mathrm{Err}(\rho) = \sum_k{w_k\left(\Tr(\rhohat E_k) - \frac{n_E}{N}\right)^2}.
\end{equation}
However, the (often neglected) weights $w_k$ are actually rather important, since they determine which of the conflicting observations will dominate.  To figure out what these weights should be, we are generally forced to turn to MLE anyway, and the best weighted least-squares fit is simply the argmax of a Gaussian approximation to the likelihood function.  For these reasons, linear inversion has been largely replaced by MLE.

Linear inversion nonetheless remains not only a powerful conceptual tool, but also the only closed-form tomographic protocol.  It proves that pretty good state tomography can in fact be done efficiently. This has never been in doubt -- but gate set tomography is a different kettle of fish. It is \emph{not} obviously feasible, for the likelihood function is not necessarily unimodal because event probabilities depend nonlinearly on the gate-set parameters.

We remedy this problem here by presenting a simple method for linear inversion gate set tomography (LGST), and a closed-form expression for the estimate.  Our approach makes implicit use of a Gram matrix technique similar to that used by Cyril Stark in \cite{Stark12a,Stark12b,Stark12c}.  We do not propose raw LGST as a final estimator -- it is clearly suboptimal in accuracy -- but as a critical part of a toolchain.  It (1) proves in principle that efficient closed-form GST is possible, and (2) provides in practice a good starting point for gradient-based likelihood maximization.

\subsection{Derivation of the LGST algorithm}

Let $\{F_k\}_{k=1\ldots d^2}$ be a set of quantum operations, each implemented by a short ``fiducial'' gate string:
\begin{equation}
F_k = G_{f_k(L)}\circ G_{f_k(L-1)}\circ\ldots G_{f_k(1)}
\end{equation}
Now, using the fixed and unknown state $\rho$ and effect $E$, let us define
\begin{eqnarray}
\sket{\rho_k} &=& F_k\sket{\rho} \nonumber \\
\sbra{E_k} &=& \sbra{E}F_k,
\end{eqnarray}
and, in terms of them, the (unknown) matrices
\begin{eqnarray}
A &=& \sum_j{\sketbra{j}{E_j}} \nonumber \\
B &=& \sum_k{\sketbra{\rho_k}{k}}.
\end{eqnarray}
Next, for any gate $X$, define
\begin{eqnarray}
\tilde{X}_{jk} &=& \sbraopket{E_j}{X}{\rho_k} \nonumber \\
&=& \sbraopket{E}{F_j X F_k}{\rho}
\end{eqnarray}
Since $\tilde{X}_{jk}$ is the probability of an experimentally observable event corresponding to the sequence $F_j X F_k$, we can ``measure'' $\tilde{X}_{jk}$ to whatever accuracy we desire, and construct the matrix
\begin{equation}
\tilde{X} = \sum_{j,k}{\sketbra{j}{k}\tilde{X}_{jk}}.
\end{equation}
Now, although we do not know the matrices $A$ and $B$, we observe that 
\begin{equation}
\tilde{X} = AXB,
\end{equation}
and in particular
\begin{equation}
\tilde{\Id} = AB.
\end{equation}
The final, critical observation is that if $\tilde{\Id}$ is invertible, then $\tilde{\Id}^{-1} = B^{-1} A^{-1}$ and
\begin{equation}
\tilde{\Id}^{-1}\tilde{X} = B^{-1}A^{-1}AXB = B^{-1}XB.
\end{equation}
So, for each gate $G_i$, we define
\begin{equation}
\hat{G}_i = \tilde{\Id}^{-1}\tilde{G}_i.
\end{equation}
This is (ignoring statistical fluctuations) a perfectly good estimate, since $\hat{G} = \{B^{-1} G_i B\}$ is gauge-equivalent to $G = \{G_i\}$.  To estimate $\sket{\rho}$ and $\sbra{E}$, we define the (element-wise identical) vectors
\begin{eqnarray}
\sket{\tilde{\rho}} &=& A\sket\rho = \sum_j{\sket{j}\sbraopket{E}{F_j}{\rho}} \\
\sbra{\tilde{E}} &=& \sbra{E}B = \sum_k{\sbraopket{E}{F_k}{\rho}\sbra{k}},
\end{eqnarray}
and observe that that linear-inversion estimates in the same gauge as the $\hat{G}_k$ estimates can be obtained as
\begin{eqnarray}
\sket{\hat{\rho}} &=& \tilde{\Id}^{-1}\sket{\tilde{\rho}} = B^{-1}\sket\rho \\
\sbra{\hat{E}} &=& \sbra{\tilde{E}} = \sbra{E}B.
\end{eqnarray}

\subsection{How to implement LGST}

The procedure for LGST is therefore to repeatedly perform each of the experiments
\begin{eqnarray*}
&\sbraopket{E}{F_j \circ G_i \circ F_k}{\rho},&\\
&\sbraopket{E}{F_j \circ F_k}{\rho},&\\
&\sbraopket{E}{F_j}{\rho},&
\end{eqnarray*}
gather statistics to estimate their probabilities, arrange those probabilities into matrices as
\begin{eqnarray}
\tilde{\Id} &=& \sum_{j,k}\sbraopket{E}{F_jF_k}{\rho}\sketbra{j}{k},\\
\tilde{G}_i &=& \sum_{j,k}\sbraopket{E}{F_jG_iF_k}{\rho}\sketbra{j}{k},\\
\sket{\tilde{\rho}} &=& \sum_j{\sbraopket{E}{F_j}{\rho}\sket{j}} \\
\sbra{\tilde{E}} &=& \sum_k{\sbraopket{E}{F_k}{\rho}\sbra{k}},
\end{eqnarray}
and then construct $\{\sket{\hat{\rho}},\hat{\sbra{E}},\{\hat{G_i}\}\}$ as above.

$\tilde{\Id}$ may not be invertible -- but if and only if either $A$ or $B$ is rank-deficient.  This occurs only if either the set $\{\rho_k\}$ or the set $\{E_j\}$ fails to span $B(\cH)$ -- i.e., they are informationally incomplete.  This is easily diagnosed by simply checking the rank of $\tilde{\Id}$.  If it occurs, we replace some of the $\{F_k\}$ with alternative sequences that \emph{do} produce informationally complete sets.  If this consistently fails to fix the rank-deficiency, it indicates that the gate set is not sufficiently universal to generate an informationally complete set, which requires hardware-level intervention.

If $\tilde{\Id}$ is full-rank, but has small eigenvalues, the experiments are marginally informationally complete.  Small statistical fluctuations in the observed frequencies will be amplified by the inversion.  This too can be fixed by adding more sequences $\{F_k\}$ to the mix and casting out the least useful ones [i.e., the ones whose removal maximizes $\lambda_{\mathrm{min}}(\tilde{\Id})$].


\section{Implementation of GST on a trapped-ion qubit} \label{sec:Experiment}

\begin{figure}[t]
\includegraphics[width=0.8 \columnwidth]{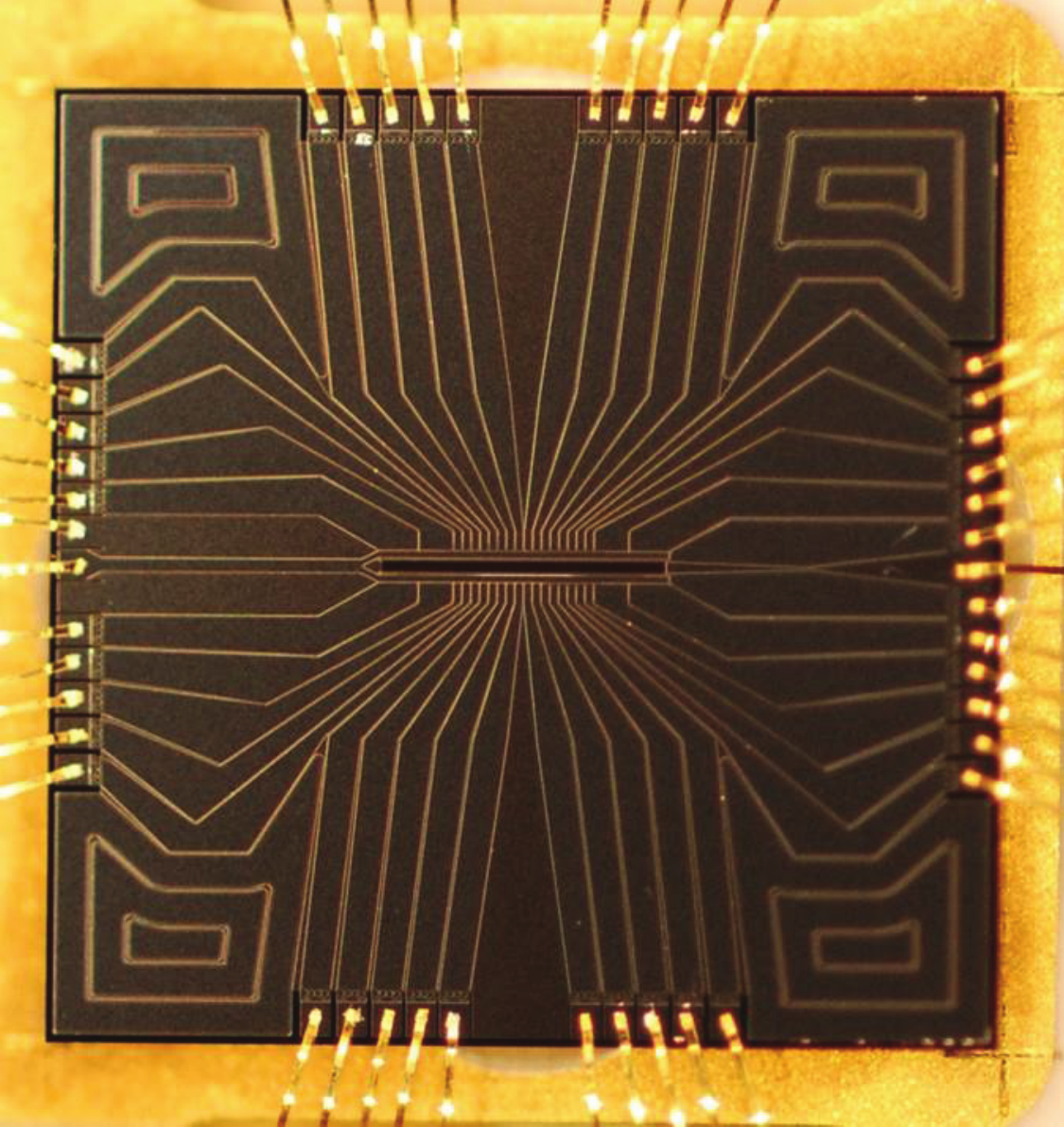}
\caption{\textbf{Surface Electrode Trap.}  Linear surface electrode ion trap used for the experiment. The trap has a central through substrate slot. Neutral ytterbium vapor reaches the trapping volume through the slot from the back of the chip. For these experiments a single {$^{171}$Yb$^+$} is trapped in the center of the trap.} \label{fig:thunderbird}
\end{figure}

Trapped ions are among the most reliable qubits available today; up to 14 qubits have been addressed in a single trap \cite{MonzPRL11}, while logic gates on single qubits have been performed with sustained failure probabilities of around $2 \times 10^{-5}$ \cite{BrownPRA11}. In order to scale these demonstrations to the large number of qubits needed for quantum information processing protocols it is crucial to use micro-fabricated trap structures. Micro-fabrication enables the fabrication of extended segmented traps that provide the ability to use multiple trapping sites and to shuttle ions between different locations.  Sandia National Laboratories uses state of the art silicon fabrication technology to produce sophisticated and highly optimized surface electrode traps for use in quantum information processing experiments.  We used a Sandia surface trap to demonstrate GST and our coherent qubit manipulation capabilities.

We trap a single $^{171}$Yb$^+$ ion in a linear surface ion trap, shown in Fig.~\ref{fig:thunderbird} \cite{stick_demonstration_2010, mount_single_2013}, by photoionizing neutral ytterbium vapor that reaches the trapping volume through a slot from the back of the surface trap chip.  The qubit is encoded in the  $| F=0, m_F=0 \rangle$ and  $| F=1, m_F=0 \rangle$ hyperfine clock states of the  $^{2}S_{1/2}$ ground state of $^{171}$Yb$^+$ which are labeled $|1\rangle$ and $|0\rangle$, respectively.

Standard laser cooling techniques are applied to Doppler cool the ion and prepare it in the $|0 \rangle$ state. The quantum state is read out via standard fluorescence state detection\cite{olmschenk_manipulation_2007}. Microwave radiation resonant with the $12.6428\operatorname{MHz}$ separation of the qubit levels is used to control the qubit. For a $\pi$-pulse microwave radiation with a square envelope is applied for approximately $58\mathrm{\mu s}$.

We used GST to characterize a set of four gates that generate the full set of single-qubit Clifford gates in this system.  Because the primary purpose of this experiment was to evaluate and demonstrate GST, we did not attempt to minimize errors in our gate set.  However, our analysis showed that the gates are extremely accurate -- enough that even standard tomography would have worked fairly well (although it is only thanks to the robust GST framework that we can say this with confidence!)

\begin{table}
\centering
\begin{tabular}{c|c|c}
 & LGST estimate ($\hat{G}_k$) & Target ($T_k$) \\
\hline
$\rho$ & $\left(\begin{array}{cc}0.0099&0.0104+0.0007i\\h.c.&0.9901\end{array}\right)$
       & $\left(\begin{array}{cc}0&0\\0&1\end{array}\right)$ \\
\hline
$E$  & $\left(\begin{array}{cc}0.9879&0.0182-0.0023i\\h.c.&0.0121\end{array}\right)$
		& $\left(\begin{array}{cc}1&0\\0&0\end{array}\right)$ \\
\hline
$G_1$ & $\left(\begin{array}{cccc}0.9977&-0.0219&-0.0204&0.0024\\-0.0152&0.9657&0.017&0.0291\\0.0031&0.0627&1.0172&0.0335\\0.001&0.0065&0.0335&0.9915\end{array}\right)$ 
	   & $\left(\begin{array}{cccc}1&0&0&0\\0&1&0&0\\0&0&1&0\\0&0&0&1\end{array}\right)$ \\
\hline
$G_2$ & $\left(\begin{array}{cccc}0.9974&-0.048&-0.0304&0.0161\\-0.0077&0.9538&-0.0033&-0.0045\\-0.0113&0.0332&0.0066&-1.0044\\-0.0029&0.0042&1.0099&0.0284\end{array}\right)$
		 & $\left(\begin{array}{cccc}1&0&0&0\\0&1&0&0\\0&0&0&-1\\0&0&1&0\end{array}\right)$ \\
\hline
$G_3$ & $\left(\begin{array}{cccc}0.9923&-0.0163&-0.0066&0.001\\-0.0049&-0.0087&-0.0087&0.9839\\0.0124&-0.0082&1.0136&-0.0017\\-0.0074&-0.9797&0.0043&0.0025\end{array}\right)$
	 	 & $\left(\begin{array}{cccc}1&0&0&0\\0&0&0&1\\0&0&1&0\\0&-1&0&0\end{array}\right)$ \\
\hline
$G_4$ & $\left(\begin{array}{cccc}0.9991&-0.0291&0.0028&0.0194\\0.0096&0.9796&-0.0049&0.0013\\0.0083&-0.0211&-1.0494&-0.0632\\-0.0091&-0.0123&-0.0427&-1.0012\end{array}\right)$
		 & $\left(\begin{array}{cccc}1&0&0&0\\0&1&0&0\\0&0&-1&0\\0&0&0&-1\end{array}\right)$ \\
\end{tabular}
\caption{\textbf{Results of LGST}.  This table shows the LGST estimate of our trapped-ion qubit gates, based on 84 distinct experiments (gate sequences), each repeated 1900 times.  The intended target gates are shown on the right.  Estimates (obtained on the left) were obtained using the LGST analysis procedure given in Section \ref{sec:LGST}, then gauge-optimized numerically, by applying similarity transformations to all gates, to match the target gates as closely as possible.} \label{tab:LGST}
\end{table}

We implemented an alphabet of four quantum operations ($\{G_1\ldots G_4\}$), aiming at the target set
\begin{eqnarray}
T_1 &=& \Id \\
T_2 &=& e^{i(\pi/4)\sigma_x} \\
T_3 &=& e^{i(\pi/4)\sigma_y} \\
T_4 &=& e^{i(\pi/2)\sigma_x}.
\end{eqnarray}
Our target initial state was $\rho_{\mathrm{ideal}} = \proj{1}$, and our target measurement was $\cM_{\mathrm{ideal}} = \{\proj{0},\proj{1}\}$.

We performed two kinds of gate sequences to gather data:  \emph{training} and \emph{testing}.  The training sequences generated data that was used to generate tomographic estimates.  The testing sequence data (discussed below) were kept hidden until after the estimates had been generated, and were then used to objectively ``score'' the four different kinds of tomography, by evaluating how well they predicted the testing data.  In order to minimize the effect of systematic drift during the several hours required to take all this data, we interleaved training and testing experiments.  We also recalibrated the gate pulse amplitude periodically.

\begin{table*}
\centering
\begin{tabular}{c|c|c|c}
 & ML estimate (short dataset) & ML estimate (long dataset) & Target gates \\
\hline
$\rho$ & $\left(\begin{array}{cc}0.0099&0.0077-0.0046i\\h.c.&0.9901\end{array}\right)$ 
		  & $\left(\begin{array}{cc}0.0092&-0.0017+0.0088i\\h.c.&0.9908\end{array}\right)$ 
		  & $\left(\begin{array}{cc}0&0\\0&1\end{array}\right)$ \\
\hline
$E$ & $\left(\begin{array}{cc}0.9911&0.0166-0.0006i\\h.c.&0.0089\end{array}\right)$
	 & $\left(\begin{array}{cc}0.988&0.0019+0.0089i\\h.c.&0.012\end{array}\right)$
	 & $\left(\begin{array}{cc}1&0\\0&0\end{array}\right)$ \\
\hline
$G_1$ & $\left(\begin{array}{cccc}1.0019&-0.0128&-0.0198&-0.0002\\-0.0066&0.9775&-0.0118&0.0122\\0.0041&0.0842&1.0138&0.0073\\-0.0035&-0.013&0.0075&0.9969\end{array}\right)$
      & $\left(\begin{array}{cccc}1.0001&-0&0.0003&0.0001\\0.0001&0.9994&-0.0003&-0\\-0.0001&0.0006&0.999&-0.0003\\-0&-0.0001&0.0002&0.9998\end{array}\right)$ 
      & $\left(\begin{array}{cccc}1&0&0&0\\0&1&0&0\\0&0&1&0\\0&0&0&1\end{array}\right)$ \\
\hline
$G_2$ & $\left(\begin{array}{cccc}1.0017&-0.0276&-0.0276&-0.0048\\-0.0193&0.9582&-0.0076&-0.0127\\-0.0134&0.043&0.0082&-0.9987\\-0.0072&0.002&1.0069&0.0192\end{array}\right)$ 
		 & $\left(\begin{array}{cccc}1&-0.0001&-0.0045&-0.0005\\0&0.9994&-0.006&-0.0018\\-0.005&-0.0112&-0.0064&-0.9991\\0.0006&0.0063&0.9993&0.0143\end{array}\right)$
		 & $\left(\begin{array}{cccc}1&0&0&0\\0&1&0&0\\0&0&0&-1\\0&0&1&0\end{array}\right)$ \\
\hline
$G_3$ & $\left(\begin{array}{cccc}0.99&-0.0114&0.0083&0.0044\\-0.0082&-0.0141&-0.0045&0.9892\\0.0121&-0.0044&1.0056&-0.0059\\-0.0001&-0.9848&0.0017&-0.0016\end{array}\right)$
		 & $\left(\begin{array}{cccc}1.0001&0.0033&0.0001&0.0049\\0.0033&-0.0001&-0.0005&0.9992\\-0.0002&-0.0024&0.9995&-0.0161\\-0.0019&-0.9989&0.0179&0.0085\end{array}\right)$
		 & $\left(\begin{array}{cccc}1&0&0&0\\0&0&0&1\\0&0&1&0\\0&-1&0&0\end{array}\right)$ \\
\hline
$G_4$ & $\left(\begin{array}{cccc}0.9983&-0.0217&0.0127&0.0142\\-0.0039&0.9745&0.0034&0.0077\\-0.0004&-0.0145&-1.0473&-0.0323\\-0.014&-0.0167&-0.0072&-1.0024\end{array}\right)$ 
		 & $\left(\begin{array}{cccc}1.0001&-0&0.0062&0.0028\\-0&0.9997&0.0127&0.0022\\0.0066&0.0164&-0.9976&0.0065\\-0.004&-0.0004&-0.0066&-0.9981\end{array}\right)$
		 & $\left(\begin{array}{cccc}1&0&0&0\\0&1&0&0\\0&0&-1&0\\0&0&0&-1\end{array}\right)$ \\
\end{tabular}
\caption{\textbf{Maximum likelihood refinements of LGST gates}.  This table shows maximum likelihood (ML) estimates of the gate set.  Column 2 shows the results of ML estimation on the ``short'' dataset of 85 sequences (the LGST sequences and the SPAM sequence $\sbraket{E}{\rho}$).  Column 3 shows the results of ML estimation on the ``long'' dataset of 1066 sequences described in the text.  Column 4 shows the target gates that we intended to implement.} \label{tab:ML}
\end{table*}

Our training sequences were designed around the demands of LGST.  We chose the simplest possible fiducial sequences, $F_k = G_k$ for $k=1\ldots 4$.  For each of the five operations $X\in\{\Id,G_1\ldots G_4\}$, we performed 16 distinct experiments of the form
$$\sbraopket{E}{F_i X F_j}{\rho}$$
to estimate the $4\times 4$ matrix $\tilde{X}$.  (For $X=\Id$, the experiments were of the form $\sbraopket{E}{F_i F_j}{\rho}$).  An additional 4 experiments involving just one gate -- $\sbraopket{E}{F_i}{\rho}$ -- were performed to enable inference of $\rho$ and $E$.  Each of these 84 experiments was repeated 1900 times to obtain statistics (each repetition yielded a single binary result, depending on whether the ion fluoresced).  The formulae from Sec. \ref{sec:LGST} were then used to calculate LGST estimates of $\rho$, $E$, and $\hat{G}_1\ldots \hat{G}_4$.  Since these estimates are only defined up to a gauge (see Sec. \ref{sec:Gauge}), we then used a numerical search to find the gauge transformation
$$\hat{G}_k \to M G_k M^{-1}$$
that minimized the RMS discrepancy,
$$\sum_k{\Tr[(G_k-T_k)^2]},$$
between the estimated $\hat{G}_k$ and the target gates $T_k$.  The resulting \emph{linear} GST estimates, represented as $4\times 4$ superoperators in the Pauli basis, and presented adjacent to the target gates, are given in Table \ref{tab:LGST}.

\section{Improving GST with maximum likelihood} \label{sec:ML}

Linear inversion tomography has been largely superseded by maximum likelihood estimation (MLE), for multiple reasons.  Linear inversion and MLE coincide when the data are informationally complete (rather than overcomplete) and the linear inversion estimate doesn't violate positivity constraints.  But the ML method can easily take account of constraints \emph{and} can reconcile overcomplete data efficiently, both of which are essentially impossible for linear inversion (least squares optimization is properly seen as an approximation to MLE, rather than a generalization of linear inversion).

For gate-set tomography, a third quality is even more important:  maximum likelihood is easily adapted to \emph{nonlinear} data -- i.e., the results of experiments in which the directly inferrable probabilities are not linear functions of the parameters.  Such experiments are natural in gate set tomography, and promise great improvements in accuracy. Probabilities involving $G_k^n$ are roughly $n$ times more sensitive to variations in $G_k$ than probabilities depending linearly on $G_k$.  However, nonlinear data poses a danger; the likelihood function $\cL(G)$ need not be unimodal or have convex level sets, which means that maximizing a generic GST likelihood function is not a convex problem, and may be NP-hard.

Fortunately, LGST provides a simple solution to this problem.  The LGST estimate is typically not optimal, but it is necessarily \emph{close} to the point of maximum likelihood, and we can reasonably expect that a gradient ascent algorithm starting from the LGST estimate will find the global maximum of $\cL(G)$.  So ML acts as a turbocharger for GST, relying critically on the LGST estimate to provide a good starting estimate, and then refining it to incorporate the nonlinear and overcomplete data that are critical for achieving high accuracy.

We use the standard Broyden-Fletcher-Goldfarb-Shanno (BFGS) algorithm \cite{nocedal2006numerical}, as implemented in \emph{Scipy} \cite{scipy}, to minimize the likelihood function. With no particular attention paid to optimization, run times ranged from a few seconds to several hours (depending on the complexity of the data) on a typical laptop computer. The LGST estimate often does not predict positive probabilities for the training data, which means that the loglikelihood is technically undefined when it is chosen as a starting point. To address this, we first find an in-bounds starting point by using the Nelder-Mead downhill simplex method \cite{Nelder1965} to find the nearest valid gate set (in terms of Euclidian distance). We then use this point as the initial value for our BFGS optimization routine.

\begin{figure*}
\includegraphics[width=0.95 \linewidth]{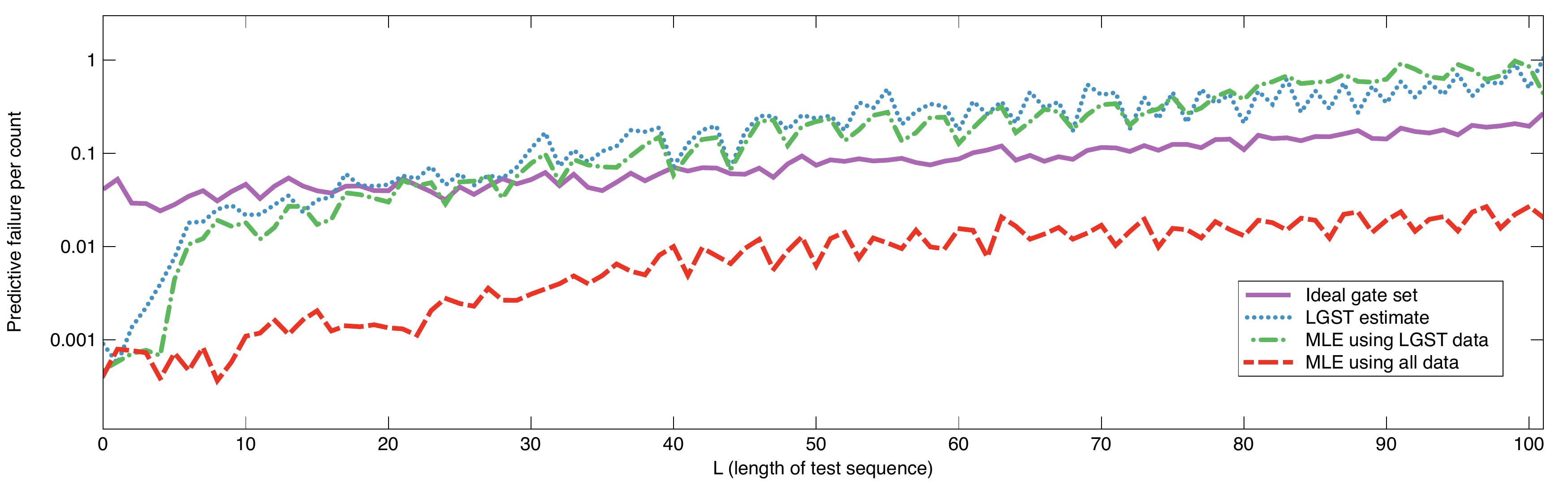}
\caption{\textbf{Average score versus test sequence length.}  This figure shows the logarithmic \emph{predictive score} achieved by four different estimates on 1000 distinct testing experiments.  Each experiment corresponded to a partial sequence comprising the first $L$ gates from one of 10 sequences of 100 gates (see Figs. \ref{fig:Rabi1}-\ref{fig:Rabi2}), and was repeated 950 times.  The vertical axis shows the average (per count) logarithmic score achieved by each of five estimates as a function of $L$ (averaged over the 10 partial sequences of length $L$).  Notably, the bare LGST estimate outperforms target gates when used to predict short sequences, but becomes rapidly very inaccurate for sequences longer than $L = 5$.  The ML estimate incorporating data from long training strings is far more accurate for $L>5$, achieving a per-count score of $\approx0.01$ even on strings of length $L=100$.  For reference, an estimate that predicted $p=\frac12$ for every experiment would suffer a score of roughly $\frac{\ln(2)}{2}\approx0.35$ per count.} \label{fig:score}
\end{figure*}


We found ML estimates for two datasets:  (1) the $84=4+16 \times 5$ LGST sequences of length $\leq3$ only, and (2) a set of $1066$ distinct sequences (each repeated 1900 times) corresponding to LGST on: $\{G_k\}$, $\{G_k^2\}$, $\{G_k^4\}$,\ldots $\{G_k^{128}\}$.  These choices of experiments were somewhat arbitrary; we have no reason to believe that these sequences will provide better (or worse) accuracy than any other sequences of various lengths.  We refer to these as the ``short'' and ``long'' datasets.  The MLE estimates are given in Table \ref{tab:ML}.

\section{Quantifying accuracy objectively with scored tests} \label{sec:score}

Tomography is not the \emph{end} of a science experiment; it is the middle.  The tomographic estimate is a theory; it needs to be tested to determine how well it predicts further experiments.  We cannot evaluate the theory based on how well it fits past (``training'') data, since its parameters were chosen specifically to fit them.  Tomographic estimates are traditionally scored using some concept of fidelity, but this is always problematic.  First, the whole point of tomography is to characterize unknown quantities, so we don't have a ``true'' state/process with which to evaluate fidelity.  Second, the gauge degree of freedom in gate set tomography makes it unclear how to calculate or interpret standard quantities like entanglement fidelity or diamond norm.

We therefore introduce a novel and very simple method for evaluating tomographic estimates.  We perform a set of ``testing'' experiments -- sequences of gates that were not performed in the tomographic phase -- and score our tomographic estimates based on how well they predict the results.  The scoring is based entirely on observable probabilities, which are explicitly gauge-invariant.  Of course, there are many ways to compare (predicted) probabilities to (empirical) frequencies.  The \emph{log scoring rule} is particularly simple and well-motivated.

\begin{figure*}
\includegraphics[width=1.0 \linewidth]{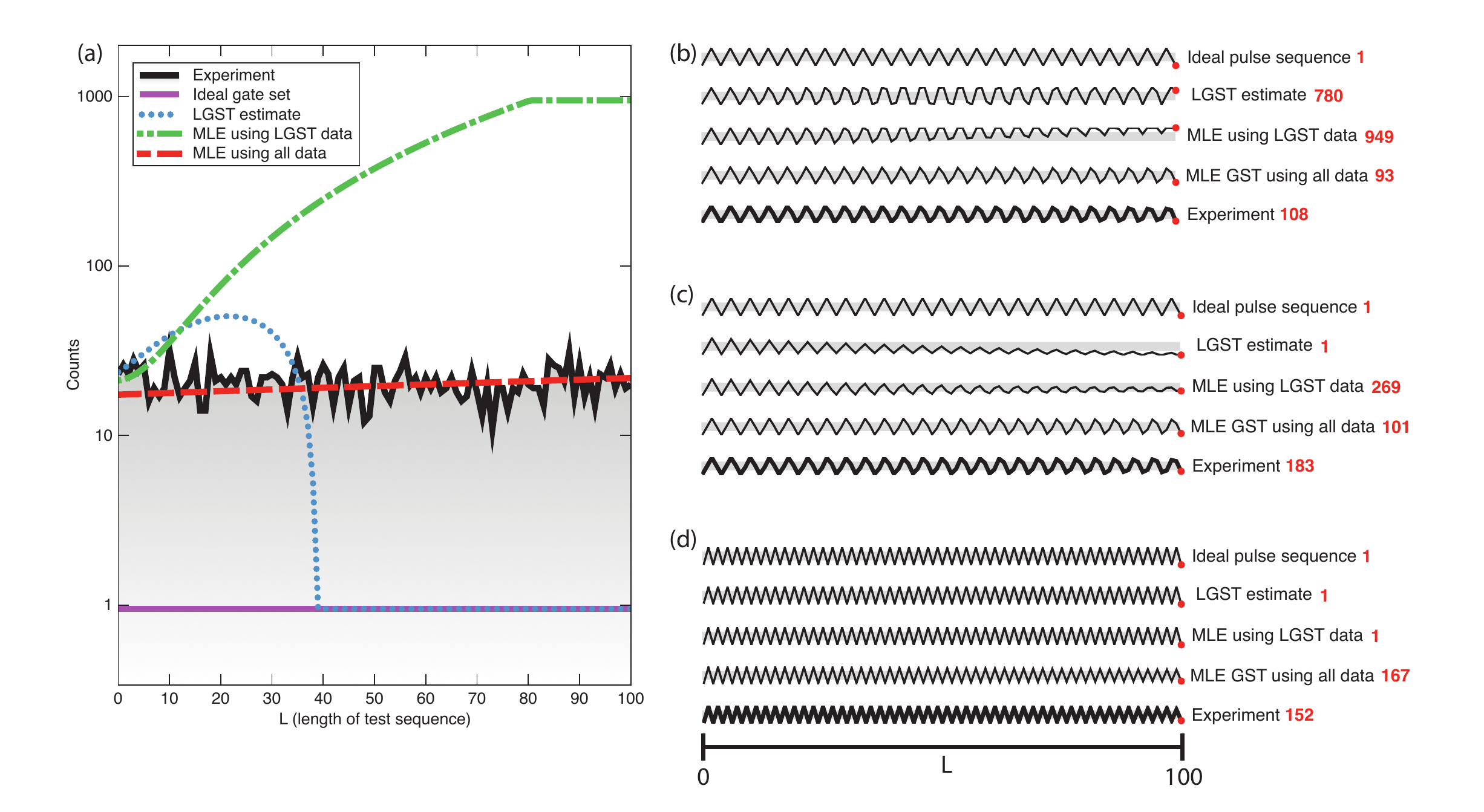}
\caption{\textbf{Rabi oscillations -- prediction vs. data}.  Each panel shows (i) observed counts and (ii) predicted counts, for one of ten series of testing experiments.  Each testing series was based on a particular sequence of 100 consecutive gates, which was used to define a series of 101 experiments corresponding to \emph{partial} sequences of length $L$ (horizontal axes).  Each partial sequence was repeated 950 times to obtain the displayed count statistics (vertical axes).  The sequences shown in this figure are: \textbf{(a)} 100 consecutive $\Id$ gates; \textbf{(b)} 100 consecutive $X_{\pi/2}$ gates; \textbf{(c)} 100 consecutive $Y_{\pi/2}$ gates; and \textbf{(d)} 100 consecutive $X_{\pi}$ gates.  Each plot thus represents a Rabi oscillation experiment (albeit with discrete gates rather than continuous Hamiltonian evolution), and compares the observed counts to the predictions of ML GST estimates obtained from both short and long training datasets. In panels \textbf{(b)-(d)}, the number after each label indicates the counts at the point at the end of each trace. The grey shaded bars represent the middle 50\%. of possible counts.} \label{fig:Rabi1}
\end{figure*}

\begin{figure*}
\includegraphics[width=1.0 \linewidth]{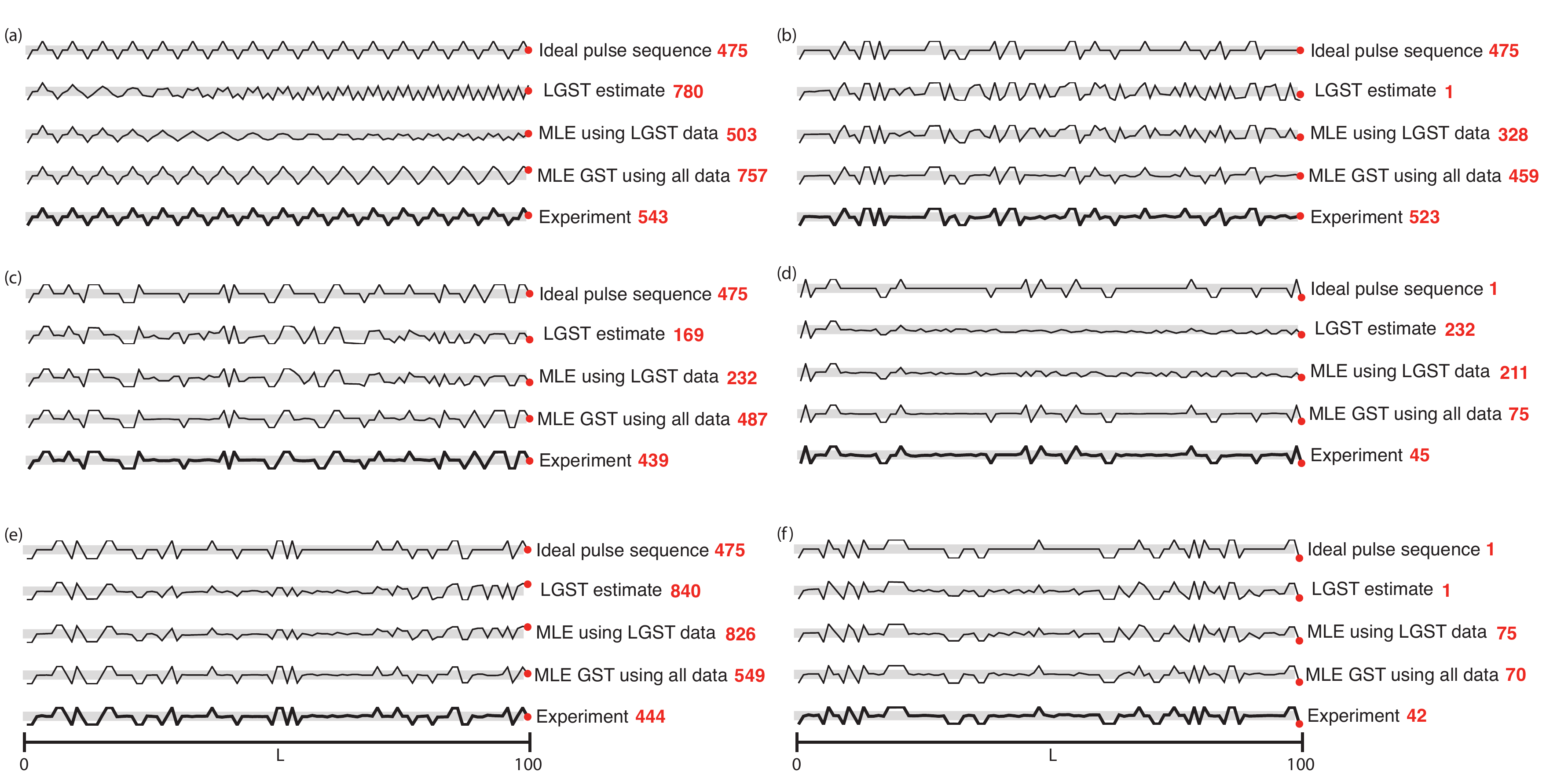}
\caption{\textbf{Generalized Rabi oscillations -- prediction vs. data}.  These plots show essentially the same sort of results as Figure \ref{fig:Rabi1}, but on non-uniform strings of gates.  Panel \textbf{(a)} shows a sequence of 100 alternating $X_{\pi/2}$ and $Y_{\pi/2}$ gates, while Panels \textbf{(b-f)} show the results of five randomly chosen gate sequences (similar to those that would be involved in randomized benchmarking).  Each plot thus represents a ``generalized Rabi oscillation experiment'', in that we are evaluating the accuracy with which a given estimate predicts an evolution through quantum state space, but that evolution is not the orbit of a Hamiltonian. In each panel, he number after each label indicates the counts at the point at the end of each trace. The grey shaded bars represent the middle 50\%. of possible counts.} \label{fig:Rabi2}
\end{figure*}

For each testing experiment $S_j$, we use the tomographic estimate[s] to assign probabilities to the outcome (before it is revealed).  Each of our experiments has 2 outcomes, which we label $+$ and $-$, so the predicted probabilities are $\{p_+, p_-\}$.  When the outcome (call it ``$b$'') is revealed, we increment the estimate's score by the negative logarithm of $Pr(b)$:
$$\mathrm{score} \to \mathrm{score} - \log(p_b).$$
When all the testing data are evaluated, this leads to a total score of
$$\mathrm{score}_0 = -\sum_j{n_+(j)\log[p_+(j)] + n_-(j)\log[p_-(j)]},$$
where lower scores are better.  We then renormalize the score by subtracting off the minimum score that \emph{any} prediction could conceivably achieve (because some of the score is due to the entropy of the data itself):
\begin{eqnarray*}
\mathrm{score} &=& \sum_j{n_+(j)\log\left[\frac{n_+(j)}{n_+(j)+n_-(j)}\right]}\\
 &+& \sum_j{n_-(j)\log\left[\frac{n_-(j)}{n_+(j)+n_-(j)}\right]}\\
 &-&\sum_j{n_+(j)\log[p_+(j)] + n_-(j)\log[p_-(j)]}.
\end{eqnarray*}
This score is in fact (1) the relative entropy between the predicted probabilities and the empirical frequencies, and (2) the loglikelihood of the tomographic estimate given the testing data.

However, while the logarithmic score is very well-motivated, it penalizes nonpositive probability estimates rather dramatically -- if $p=0$, then the penalty is $-\log(0)=\infty$, and if $p<0$, the whole formalism fails.  Although negative probabilities should never be predicted, we did not impose positivity in this work, so some of our estimates (especially naive tomography and LGST) do predict negative probabilities.  We deal with this in a simple and tolerably well-motivated way:  whenever an estimate predicts $p<\epsilon$ for some small threshold $\epsilon$, we truncate that probability to $\epsilon$.  In the data reported here, we chose $\epsilon=10^{-3}$ [approximately $1/N$, where each training sequence was measured $N=1900$ times; after $N$ observations, the lowest probability that can reasonably be reported is $O(1/N)$], and verified that varying $\epsilon$ does not qualitatively change the results.

We scored and compared four different estimates, each of which can be used to predict the testing data:
\begin{enumerate}
\item The target gates themselves,
\item LGST using only the $4+5\times16 = 84$ training sequences discussed above,
\item Maximum likelihood GST on the ``short'' 85-sequence dataset comprising 84 LGST sequences and the ``SPAM'' experiment $\sbraket{E}{\rho}$,
\item Maximum likelihood GST using a much richer ``long'' dataset, described below.
\end{enumerate}

Our ``long'' dataset was intended to probe the use of long gate sequences: (1) their utility for improving accuracy; and (2) the feasibility of GST estimation for such data.  We included the 85 sequences in the ``short'' dataset, and added $448 = 7 \times 4 \times 16$ additional sequences of the form $\sbraopket{E}{F_i G_k^p F_j}{\rho}$ for $p=2,4,8,16,32,64,128$.  That is, we did the experiments necessary for LGST on $\{G_k^p\}$, although we only analyzed this data using MLE.  Finally, for each of these 533 experiments, we \emph{also} performed a corresponding experiment in which we added a single $G_4 \approx e^{i(\pi/2)\sigma_x}$ gate at the end, so that we could probe the bright-vs-dark asymmetry of the measurement\footnote{This turned out to not actually be necessary -- GST is completely self-calibrating, and can extract this asymmetry from any complete set of training sequences.}.  The ``long'' dataset thus contains a total of 1066 sequences, each repeated 1900 times, for a total of $\approx2\times10^6$ measurements. 

The main results are displayed in Figure \ref{fig:score}, which shows the estimates' score-per-count (note: lower scores are better), averaged over all 10 test sequences of length $L$, as a function of $L$.  Shorter sequences are easier to predict, and all estimates' scores increase with $L$.

All three tomographic estimates predict very short ($L\leq5$) fairly well. The target gates themselves fail to predict even short sequences well, although most of this predictive failure seems to be due to SPAM error rather than errors in the gates.  Maximum likelihood methods are more accurate than LGST even for strings of length $L=4,5$, but the most dramatic difference is between the ``Long ML'' estimates, which were trained on long sequences, and all the others.  For reference, we note that an estimator that simply predicted $p=\frac12$ for every count would achieve a score-per-count of approximately $\frac{\ln2}{2}\approx0.35$.  The best ML estimate still achieves a score-per-count of $\sim0.02$ at $L=100$, indicating a very high degree of predictive power even for long test sequences.

The LGST estimate works well on strings of the same length as its training data, but degrades rapidly beyond $L=3$.  This is not a major concern, however.  LGST's critical role is to get \emph{close enough} to provide a good seed for ML methods (which it does admirably), not to provide an optimal estimate.  We suspect that LGST's accuracy could be improved quite a bit by using overcomplete data and appropriately weighted least-squares fitting.

\section{Conclusions} \label{sec:Conclusions}

Continued development of QIP technology -- memory qubits, logic gates, state preparations, and measurements -- depends critically on reliable characterization protocols, which cannot rely on precalibrated reference frames that are not available in most technologies.  Gate set tomography is, to our knowledge, the first completely reliable framework and protocol for characterizing quantum logic gates.  By using LGST as a first stage to obtain a closed-form approximation to the true gates, the GST protocol can ensure robustness against local maxima in the likelihood function -- yet take full advantage of maximum likelihood (or any other well-motivated statistical method) to achieve high accuracy.

Our experimental demonstration illustrates GST's ability, \emph{and} demonstrates that it is practically feasible.  Moreover, out of necessity, we have introduced and demonstrated a novel and (we think) very useful method for objectively testing how good a tomographic estimate is.  Unlike all the previous work of which we are aware, this scoring protocol doesn't measure how well the tomographic estimate agrees with the target goal (which might be incorrect) or with another tomographic estimate (which might be biased in the same way as this one).  Instead, it evaluates how well the estimate does its fundamental job -- predicting future data.  Our results not only illustrate the scoring protocol, but also show that our GST estimates are quite good predictors.

We do not expect that gate set tomography will be another kind of tomography, standing shoulder to shoulder with state tomography, process tomography, and measurement tomography.  It is intended to \emph{replace} them -- to be, as one of us has said in public, ``One tomography to rule them all.''  This is out of necessity:  state tomography requires well-calibrated measurements, measurement tomography requires well-calibrated states, and gate tomography requires both -- yet in practice, states and measurements are only as well calibrated as the gates that prepare them!  This is a vicious circle.  GST cuts that Gordian knot by (1) estimating everything self-consistently, and (2) identifying the \emph{gates} as the critical element.  Gates are central because they can be applied multiple times in a single experiment (unlike state preparations and measurements, which can appear only once per experiment), and this allows us to generate combinatorially many ($2^L$) distinct observable probabilities using only 2 distinct gates (and thus without adding any extra parameters to be estimated).

The necessary price paid for this is the appearance of the $SL(d^2)$ gauge.  This gauge is arguably the single most intriguing and pernicious aspect of GST.  It is clearly fundamental to black-box descriptions of quantum devices, and therefore seems to be fundamental to QIP.  Yet it interacts very badly with complete positivity, and we do not yet know how to represent gate sets in an efficient and gauge-invariant way -- nor how to compute a gauge-invariant measure of fidelity between two gate sets (e.g., a target and an estimate).  Of all the many aspects of GST that cry out for further research and development, the gauge -- its relationship with conventional descriptions of circuit QIP, and how it can be tamed -- seems the most worthy of urgent study.

\begin{acknowledgments}
RBK is grateful for helpful conversations with Steve Flammia, Stephen Bartlett, Jay Gambetta, Seth Merkel, and Cyril Stark.  Sandia National Laboratories is a multi-program laboratory operated by Sandia Corporation, a wholly owned subsidiary of Lockheed Martin Corporation, for the U.S. Department of Energy's National Nuclear Security Administration under contract DE-AC04-94AL85000. 
\end{acknowledgments}

\bibliography{LGST}{}

\begin{thebibliography}{29}
\expandafter\ifx\csname natexlab\endcsname\relax\def\natexlab#1{#1}\fi
\expandafter\ifx\csname bibnamefont\endcsname\relax
  \def\bibnamefont#1{#1}\fi
\expandafter\ifx\csname bibfnamefont\endcsname\relax
  \def\bibfnamefont#1{#1}\fi
\expandafter\ifx\csname citenamefont\endcsname\relax
  \def\citenamefont#1{#1}\fi
\expandafter\ifx\csname url\endcsname\relax
  \def\url#1{\texttt{#1}}\fi
\expandafter\ifx\csname urlprefix\endcsname\relax\def\urlprefix{URL }\fi
\providecommand{\bibinfo}[2]{#2}
\providecommand{\eprint}[2][]{\url{#2}}

\bibitem[{\citenamefont{Merkel et~al.}(2013)\citenamefont{Merkel, Gambetta,
  Smolin, Poletto, C\'orcoles, Johnson, Ryan, and Steffen}}]{MerkelPRA13}
\bibinfo{author}{\bibfnamefont{S.~T.} \bibnamefont{Merkel}},
  \bibinfo{author}{\bibfnamefont{J.~M.} \bibnamefont{Gambetta}},
  \bibinfo{author}{\bibfnamefont{J.~A.} \bibnamefont{Smolin}},
  \bibinfo{author}{\bibfnamefont{S.}~\bibnamefont{Poletto}},
  \bibinfo{author}{\bibfnamefont{A.~D.} \bibnamefont{C\'orcoles}},
  \bibinfo{author}{\bibfnamefont{B.~R.} \bibnamefont{Johnson}},
  \bibinfo{author}{\bibfnamefont{C.~A.} \bibnamefont{Ryan}}, \bibnamefont{and}
  \bibinfo{author}{\bibfnamefont{M.}~\bibnamefont{Steffen}},
  \bibinfo{journal}{Phys. Rev. A} \textbf{\bibinfo{volume}{87}},
  \bibinfo{pages}{062119} (\bibinfo{year}{2013}).

\bibitem[{\citenamefont{Takahashi et~al.}(2013)\citenamefont{Takahashi,
  Bartlett, and Doherty}}]{TakahashiPRA13}
\bibinfo{author}{\bibfnamefont{M.}~\bibnamefont{Takahashi}},
  \bibinfo{author}{\bibfnamefont{S.~D.} \bibnamefont{Bartlett}},
  \bibnamefont{and} \bibinfo{author}{\bibfnamefont{A.~C.}
  \bibnamefont{Doherty}}, \bibinfo{journal}{Phys. Rev. A}
  \textbf{\bibinfo{volume}{88}}, \bibinfo{pages}{022120}
  (\bibinfo{year}{2013}).

\bibitem[{\citenamefont{{Medford} et~al.}(2013)\citenamefont{{Medford}, {Beil},
  {Taylor}, {Bartlett}, {Doherty}, {Rashba}, {Divincenzo}, {Lu}, {Gossard}, and
  {Marcus}}}]{MedfordNN13}
\bibinfo{author}{\bibfnamefont{J.}~\bibnamefont{{Medford}}},
  \bibinfo{author}{\bibfnamefont{J.}~\bibnamefont{{Beil}}},
  \bibinfo{author}{\bibfnamefont{J.~M.} \bibnamefont{{Taylor}}},
  \bibinfo{author}{\bibfnamefont{S.~D.} \bibnamefont{{Bartlett}}},
  \bibinfo{author}{\bibfnamefont{A.~C.} \bibnamefont{{Doherty}}},
  \bibinfo{author}{\bibfnamefont{E.~I.} \bibnamefont{{Rashba}}},
  \bibinfo{author}{\bibfnamefont{D.~P.} \bibnamefont{{Divincenzo}}},
  \bibinfo{author}{\bibfnamefont{H.}~\bibnamefont{{Lu}}},
  \bibinfo{author}{\bibfnamefont{A.~C.} \bibnamefont{{Gossard}}},
  \bibnamefont{and} \bibinfo{author}{\bibfnamefont{C.~M.}
  \bibnamefont{{Marcus}}}, \bibinfo{journal}{Nature Nanotechnology}
  \textbf{\bibinfo{volume}{8}}, \bibinfo{pages}{654} (\bibinfo{year}{2013}).

\bibitem[{\citenamefont{Mahler et~al.}(2013)\citenamefont{Mahler, Rozema,
  Darabi, Ferrie, Blume-Kohout, and Steinberg}}]{MahlerPRL13}
\bibinfo{author}{\bibfnamefont{D.}~\bibnamefont{Mahler}},
  \bibinfo{author}{\bibfnamefont{L.~A.} \bibnamefont{Rozema}},
  \bibinfo{author}{\bibfnamefont{A.}~\bibnamefont{Darabi}},
  \bibinfo{author}{\bibfnamefont{C.}~\bibnamefont{Ferrie}},
  \bibinfo{author}{\bibfnamefont{R.}~\bibnamefont{Blume-Kohout}},
  \bibnamefont{and}
  \bibinfo{author}{\bibfnamefont{A.}~\bibnamefont{Steinberg}},
  \bibinfo{journal}{arXiv preprint arXiv:1303.0436}  (\bibinfo{year}{2013}).

\bibitem[{\citenamefont{Brown et~al.}(2011)\citenamefont{Brown, Wilson,
  Colombe, Ospelkaus, Meier, Knill, Leibfried, and Wineland}}]{BrownPRA11}
\bibinfo{author}{\bibfnamefont{K.~R.} \bibnamefont{Brown}},
  \bibinfo{author}{\bibfnamefont{A.~C.} \bibnamefont{Wilson}},
  \bibinfo{author}{\bibfnamefont{Y.}~\bibnamefont{Colombe}},
  \bibinfo{author}{\bibfnamefont{C.}~\bibnamefont{Ospelkaus}},
  \bibinfo{author}{\bibfnamefont{A.~M.} \bibnamefont{Meier}},
  \bibinfo{author}{\bibfnamefont{E.}~\bibnamefont{Knill}},
  \bibinfo{author}{\bibfnamefont{D.}~\bibnamefont{Leibfried}},
  \bibnamefont{and} \bibinfo{author}{\bibfnamefont{D.~J.}
  \bibnamefont{Wineland}}, \bibinfo{journal}{Phys. Rev. A}
  \textbf{\bibinfo{volume}{84}}, \bibinfo{pages}{030303}
  (\bibinfo{year}{2011}).

\bibitem[{\citenamefont{Mogilevtsev et~al.}(2012)\citenamefont{Mogilevtsev,
  {\v{R}}eh{\'a}{\v{c}}ek, and Hradil}}]{MogilevtsevNJP12}
\bibinfo{author}{\bibfnamefont{D.}~\bibnamefont{Mogilevtsev}},
  \bibinfo{author}{\bibfnamefont{J.}~\bibnamefont{{\v{R}}eh{\'a}{\v{c}}ek}},
  \bibnamefont{and} \bibinfo{author}{\bibfnamefont{Z.}~\bibnamefont{Hradil}},
  \bibinfo{journal}{New Journal of Physics} \textbf{\bibinfo{volume}{14}},
  \bibinfo{pages}{095001} (\bibinfo{year}{2012}).

\bibitem[{\citenamefont{Bra{\'n}czyk et~al.}(2012)\citenamefont{Bra{\'n}czyk,
  Mahler, Rozema, Darabi, Steinberg, and James}}]{BranczykNJP12}
\bibinfo{author}{\bibfnamefont{A.}~\bibnamefont{Bra{\'n}czyk}},
  \bibinfo{author}{\bibfnamefont{D.~H.} \bibnamefont{Mahler}},
  \bibinfo{author}{\bibfnamefont{L.~A.} \bibnamefont{Rozema}},
  \bibinfo{author}{\bibfnamefont{A.}~\bibnamefont{Darabi}},
  \bibinfo{author}{\bibfnamefont{A.~M.} \bibnamefont{Steinberg}},
  \bibnamefont{and} \bibinfo{author}{\bibfnamefont{D.~F.} \bibnamefont{James}},
  \bibinfo{journal}{New Journal of Physics} \textbf{\bibinfo{volume}{14}},
  \bibinfo{pages}{085003} (\bibinfo{year}{2012}).

\bibitem[{\citenamefont{Stark}(2012{\natexlab{a}})}]{Stark12a}
\bibinfo{author}{\bibfnamefont{C.}~\bibnamefont{Stark}},
  \bibinfo{journal}{arXiv preprint arXiv:1209.5737}
  (\bibinfo{year}{2012}{\natexlab{a}}).

\bibitem[{\citenamefont{Stark}(2012{\natexlab{b}})}]{Stark12b}
\bibinfo{author}{\bibfnamefont{C.}~\bibnamefont{Stark}},
  \bibinfo{journal}{arXiv preprint arXiv:1209.6499}
  (\bibinfo{year}{2012}{\natexlab{b}}).

\bibitem[{\citenamefont{Stark}(2012{\natexlab{c}})}]{Stark12c}
\bibinfo{author}{\bibfnamefont{C.}~\bibnamefont{Stark}},
  \bibinfo{journal}{arXiv preprint arXiv:1210.1105}
  (\bibinfo{year}{2012}{\natexlab{c}}).

\bibitem[{\citenamefont{Vogel and Risken}(1989)}]{VogelPRA89}
\bibinfo{author}{\bibfnamefont{K.}~\bibnamefont{Vogel}} \bibnamefont{and}
  \bibinfo{author}{\bibfnamefont{H.}~\bibnamefont{Risken}},
  \bibinfo{journal}{Phys. Rev. A} \textbf{\bibinfo{volume}{40}},
  \bibinfo{pages}{2847} (\bibinfo{year}{1989}).

\bibitem[{\citenamefont{Smithey et~al.}(1993)\citenamefont{Smithey, Beck,
  Raymer, and Faridani}}]{SmitheyPRL93}
\bibinfo{author}{\bibfnamefont{D.}~\bibnamefont{Smithey}},
  \bibinfo{author}{\bibfnamefont{M.}~\bibnamefont{Beck}},
  \bibinfo{author}{\bibfnamefont{M.}~\bibnamefont{Raymer}}, \bibnamefont{and}
  \bibinfo{author}{\bibfnamefont{A.}~\bibnamefont{Faridani}},
  \bibinfo{journal}{Physical review letters} \textbf{\bibinfo{volume}{70}},
  \bibinfo{pages}{1244} (\bibinfo{year}{1993}).

\bibitem[{\citenamefont{Paris and {\v{R}}eh{\'a}{\v{c}}ek}(2004)}]{ParisBook04}
\bibinfo{author}{\bibfnamefont{M.}~\bibnamefont{Paris}} \bibnamefont{and}
  \bibinfo{author}{\bibfnamefont{J.}~\bibnamefont{{\v{R}}eh{\'a}{\v{c}}ek}},
  \emph{\bibinfo{title}{Quantum state estimation}}, vol. \bibinfo{volume}{649}
  (\bibinfo{publisher}{Springer}, \bibinfo{year}{2004}).

\bibitem[{\citenamefont{Chuang and Nielsen}(1997)}]{ChuangJMO97}
\bibinfo{author}{\bibfnamefont{I.~L.} \bibnamefont{Chuang}} \bibnamefont{and}
  \bibinfo{author}{\bibfnamefont{M.}~\bibnamefont{Nielsen}},
  \bibinfo{journal}{Journal of Modern Optics} \textbf{\bibinfo{volume}{44}},
  \bibinfo{pages}{2455} (\bibinfo{year}{1997}).

\bibitem[{\citenamefont{Hradil}(1997)}]{HradilPRA97}
\bibinfo{author}{\bibfnamefont{Z.}~\bibnamefont{Hradil}},
  \bibinfo{journal}{Physical Review A} \textbf{\bibinfo{volume}{55}},
  \bibinfo{pages}{1561} (\bibinfo{year}{1997}).

\bibitem[{\citenamefont{Blume-Kohout}(2010{\natexlab{a}})}]{RBKNJP10}
\bibinfo{author}{\bibfnamefont{R.}~\bibnamefont{Blume-Kohout}},
  \bibinfo{journal}{New J. Phys.} \textbf{\bibinfo{volume}{12}},
  \bibinfo{pages}{043034} (\bibinfo{year}{2010}{\natexlab{a}}).

\bibitem[{\citenamefont{Blume-Kohout}(2010{\natexlab{b}})}]{RBKPRL10}
\bibinfo{author}{\bibfnamefont{R.}~\bibnamefont{Blume-Kohout}},
  \bibinfo{journal}{Phys. Rev. Lett.} \textbf{\bibinfo{volume}{105}},
  \bibinfo{pages}{200504} (\bibinfo{year}{2010}{\natexlab{b}}).

\bibitem[{\citenamefont{Smolin et~al.}(2012)\citenamefont{Smolin, Gambetta, and
  Smith}}]{SmolinPRL12}
\bibinfo{author}{\bibfnamefont{J.~A.} \bibnamefont{Smolin}},
  \bibinfo{author}{\bibfnamefont{J.~M.} \bibnamefont{Gambetta}},
  \bibnamefont{and} \bibinfo{author}{\bibfnamefont{G.}~\bibnamefont{Smith}},
  \bibinfo{journal}{Phys. Rev. Lett.} \textbf{\bibinfo{volume}{108}},
  \bibinfo{pages}{070502} (\bibinfo{year}{2012}).

\bibitem[{\citenamefont{Cramer et~al.}(2010)\citenamefont{Cramer, Plenio,
  Flammia, Somma, Gross, Bartlett, Landon-Cardinal, Poulin, and
  Liu}}]{CramerNC10}
\bibinfo{author}{\bibfnamefont{M.}~\bibnamefont{Cramer}},
  \bibinfo{author}{\bibfnamefont{M.~B.} \bibnamefont{Plenio}},
  \bibinfo{author}{\bibfnamefont{S.~T.} \bibnamefont{Flammia}},
  \bibinfo{author}{\bibfnamefont{R.}~\bibnamefont{Somma}},
  \bibinfo{author}{\bibfnamefont{D.}~\bibnamefont{Gross}},
  \bibinfo{author}{\bibfnamefont{S.~D.} \bibnamefont{Bartlett}},
  \bibinfo{author}{\bibfnamefont{O.}~\bibnamefont{Landon-Cardinal}},
  \bibinfo{author}{\bibfnamefont{D.}~\bibnamefont{Poulin}}, \bibnamefont{and}
  \bibinfo{author}{\bibfnamefont{Y.-K.} \bibnamefont{Liu}},
  \bibinfo{journal}{Nature Communications} \textbf{\bibinfo{volume}{1}},
  \bibinfo{pages}{149} (\bibinfo{year}{2010}).

\bibitem[{\citenamefont{Gross et~al.}(2010)\citenamefont{Gross, Liu, Flammia,
  Becker, and Eisert}}]{GrossPRL10}
\bibinfo{author}{\bibfnamefont{D.}~\bibnamefont{Gross}},
  \bibinfo{author}{\bibfnamefont{Y.-K.} \bibnamefont{Liu}},
  \bibinfo{author}{\bibfnamefont{S.~T.} \bibnamefont{Flammia}},
  \bibinfo{author}{\bibfnamefont{S.}~\bibnamefont{Becker}}, \bibnamefont{and}
  \bibinfo{author}{\bibfnamefont{J.}~\bibnamefont{Eisert}},
  \bibinfo{journal}{Phys. Rev. Lett.} \textbf{\bibinfo{volume}{105}},
  \bibinfo{pages}{150401} (\bibinfo{year}{2010}),
  \urlprefix\url{http://link.aps.org/doi/10.1103/PhysRevLett.105.150401}.

\bibitem[{\citenamefont{van Dam et~al.}(2000)\citenamefont{van Dam, Magniez,
  Mosca, and Santha}}]{vanDamACM00}
\bibinfo{author}{\bibfnamefont{W.}~\bibnamefont{van Dam}},
  \bibinfo{author}{\bibfnamefont{F.}~\bibnamefont{Magniez}},
  \bibinfo{author}{\bibfnamefont{M.}~\bibnamefont{Mosca}}, \bibnamefont{and}
  \bibinfo{author}{\bibfnamefont{M.}~\bibnamefont{Santha}}, in
  \emph{\bibinfo{booktitle}{Proceedings of the thirty-second annual ACM
  symposium on Theory of computing}} (\bibinfo{organization}{ACM},
  \bibinfo{year}{2000}), pp. \bibinfo{pages}{688--696}.

\bibitem[{\citenamefont{Blume-Kohout and et. al.}(2013)}]{RBKinprep}
\bibinfo{author}{\bibfnamefont{R.}~\bibnamefont{Blume-Kohout}}
  \bibnamefont{and} \bibinfo{author}{\bibnamefont{et. al.}},
  \bibinfo{journal}{in preparation}  (\bibinfo{year}{2013}).

\bibitem[{\citenamefont{Monz et~al.}(2011)\citenamefont{Monz, Schindler,
  Barreiro, Chwalla, Nigg, Coish, Harlander, H{\"a}nsel, Hennrich, and
  Blatt}}]{MonzPRL11}
\bibinfo{author}{\bibfnamefont{T.}~\bibnamefont{Monz}},
  \bibinfo{author}{\bibfnamefont{P.}~\bibnamefont{Schindler}},
  \bibinfo{author}{\bibfnamefont{J.~T.} \bibnamefont{Barreiro}},
  \bibinfo{author}{\bibfnamefont{M.}~\bibnamefont{Chwalla}},
  \bibinfo{author}{\bibfnamefont{D.}~\bibnamefont{Nigg}},
  \bibinfo{author}{\bibfnamefont{W.~A.} \bibnamefont{Coish}},
  \bibinfo{author}{\bibfnamefont{M.}~\bibnamefont{Harlander}},
  \bibinfo{author}{\bibfnamefont{W.}~\bibnamefont{H{\"a}nsel}},
  \bibinfo{author}{\bibfnamefont{M.}~\bibnamefont{Hennrich}}, \bibnamefont{and}
  \bibinfo{author}{\bibfnamefont{R.}~\bibnamefont{Blatt}},
  \bibinfo{journal}{Phys. Rev. Lett.} \textbf{\bibinfo{volume}{106}},
  \bibinfo{pages}{130506} (\bibinfo{year}{2011}).

\bibitem[{\citenamefont{Stick et~al.}(2010)\citenamefont{Stick, Fortier,
  Haltli, Highstrete, Moehring, Tigges, and Blain}}]{stick_demonstration_2010}
\bibinfo{author}{\bibfnamefont{D.}~\bibnamefont{Stick}},
  \bibinfo{author}{\bibfnamefont{K.~M.} \bibnamefont{Fortier}},
  \bibinfo{author}{\bibfnamefont{R.}~\bibnamefont{Haltli}},
  \bibinfo{author}{\bibfnamefont{C.}~\bibnamefont{Highstrete}},
  \bibinfo{author}{\bibfnamefont{D.~L.} \bibnamefont{Moehring}},
  \bibinfo{author}{\bibfnamefont{C.}~\bibnamefont{Tigges}}, \bibnamefont{and}
  \bibinfo{author}{\bibfnamefont{M.~G.} \bibnamefont{Blain}},
  \bibinfo{journal}{arXiv preprint arXiv:1008.0990}  (\bibinfo{year}{2010}).

\bibitem[{\citenamefont{Mount et~al.}(2013)\citenamefont{Mount, Baek, Blain,
  Stick, Gaultney, Crain, Noek, Kim, Maunz, and Kim}}]{mount_single_2013}
\bibinfo{author}{\bibfnamefont{E.}~\bibnamefont{Mount}},
  \bibinfo{author}{\bibfnamefont{S.-Y.} \bibnamefont{Baek}},
  \bibinfo{author}{\bibfnamefont{M.}~\bibnamefont{Blain}},
  \bibinfo{author}{\bibfnamefont{D.}~\bibnamefont{Stick}},
  \bibinfo{author}{\bibfnamefont{D.}~\bibnamefont{Gaultney}},
  \bibinfo{author}{\bibfnamefont{S.}~\bibnamefont{Crain}},
  \bibinfo{author}{\bibfnamefont{R.}~\bibnamefont{Noek}},
  \bibinfo{author}{\bibfnamefont{T.}~\bibnamefont{Kim}},
  \bibinfo{author}{\bibfnamefont{P.}~\bibnamefont{Maunz}}, \bibnamefont{and}
  \bibinfo{author}{\bibfnamefont{J.}~\bibnamefont{Kim}},
  \bibinfo{journal}{arXiv preprint arXiv:1306.1269}  (\bibinfo{year}{2013}).

\bibitem[{\citenamefont{Olmschenk et~al.}(2007)\citenamefont{Olmschenk, Younge,
  Moehring, Matsukevich, Maunz, and Monroe}}]{olmschenk_manipulation_2007}
\bibinfo{author}{\bibfnamefont{S.}~\bibnamefont{Olmschenk}},
  \bibinfo{author}{\bibfnamefont{K.~C.} \bibnamefont{Younge}},
  \bibinfo{author}{\bibfnamefont{D.~L.} \bibnamefont{Moehring}},
  \bibinfo{author}{\bibfnamefont{D.~N.} \bibnamefont{Matsukevich}},
  \bibinfo{author}{\bibfnamefont{P.}~\bibnamefont{Maunz}}, \bibnamefont{and}
  \bibinfo{author}{\bibfnamefont{C.}~\bibnamefont{Monroe}},
  \bibinfo{journal}{Physical Review A} \textbf{\bibinfo{volume}{76}},
  \bibinfo{pages}{052314} (\bibinfo{year}{2007}).

\bibitem[{\citenamefont{Nocedal and Wright}(2006)}]{nocedal2006numerical}
\bibinfo{author}{\bibfnamefont{J.}~\bibnamefont{Nocedal}} \bibnamefont{and}
  \bibinfo{author}{\bibfnamefont{S.}~\bibnamefont{Wright}},
  \emph{\bibinfo{title}{Numerical optimization}}, Operations research and
  financial engineering (\bibinfo{publisher}{Springer, New York},
  \bibinfo{year}{2006}).

\bibitem[{\citenamefont{Jones et~al.}(2001--)\citenamefont{Jones, Oliphant,
  Peterson et~al.}}]{scipy}
\bibinfo{author}{\bibfnamefont{E.}~\bibnamefont{Jones}},
  \bibinfo{author}{\bibfnamefont{T.}~\bibnamefont{Oliphant}},
  \bibinfo{author}{\bibfnamefont{P.}~\bibnamefont{Peterson}},
  \bibnamefont{et~al.}, \emph{\bibinfo{title}{{SciPy}: Open source scientific
  tools for {Python}}} (\bibinfo{year}{2001--}),
  \urlprefix\url{http://www.scipy.org/}.

\bibitem[{\citenamefont{Nelder and Mead}(1965)}]{Nelder1965}
\bibinfo{author}{\bibfnamefont{J.}~\bibnamefont{Nelder}} \bibnamefont{and}
  \bibinfo{author}{\bibfnamefont{R.}~\bibnamefont{Mead}},
  \bibinfo{journal}{Computer Journal} \textbf{\bibinfo{volume}{7}},
  \bibinfo{pages}{308} (\bibinfo{year}{1965}).

\end{thebibliography}

\end{document}